\documentclass[aps,showpacs,preprintnumbers,amsmath]{revtex4-2}
\usepackage{graphicx}% Include figure files
\usepackage{dcolumn}% Align table columns on decimal point
\usepackage{epstopdf}
\usepackage{color}
\usepackage{bm}% bold math
\begin{document}
%\begin{frontmatter}
\title{Percolation and de-confinement in relativistic nuclear collisions}
\author{ B. K. Srivastava {$^1$}, R. P. Scharenberg{$^1$ }, and C. Pajares{$^2$}}
\medskip
\affiliation{$^1$Department of Physics and Astronomy, Purdue University, West Lafayette, IN-47907, USA\\ 
$^2$Departamento de Fisica de Particulas, Universidale de Santiago de Compostela and Instituto Galego de Fisica de Atlas Enerxias(IGFAE), 15782 Santiago, de Compostela, Spain}
\bigskip
    
 \begin{abstract}
 In the present work we have analyzed the transverse momentum spectra  of charged particles in high multiplicity $pp$ collisions at LHC energies 
$\sqrt s $ = 5.02 and 13  TeV using the Color String Percolation Model (CSPM).
For heavy ions  $Pb-Pb$ at $\sqrt {s_{NN}} $ = 2.76 and 5.02 TeV along with
$Xe-Xe$ at $\sqrt {s_{NN}} $= 5.44 TeV have been analyzed. The initial temperature is extracted both in low and high multiplicity events in ${\it pp}$ collisions. For $A-A$ collisions the temperature is obtained as a function of centrality.
 %A universal scaling in the temperature from $pp$ and $A-A$ collisions is obtai%ned when multiplicity is scaled by the transverse interaction area.
 From the measured energy density $ \varepsilon$ and the temperature T the dimensionless quantity $ \varepsilon/T^{4}$ is obtained. Our results for Pb-Pb and Xe-Xe collisions show a sharp increase in $\varepsilon/T^{4}$ above T $\sim$ 210 MeV and reaching the ideal gas of quarks and gluons value of $ \varepsilon/T^{4} \sim$ 16 at temperature $\sim $ 230 MeV. At this temperature there
is a transition from the fluid behavior of QCD matter strongly interacting
to a quasi free gas of quarks and gluons.
\end{abstract}
%\FullConference{53rd International Symposium on Multiparticle Dynamics,\\
% 21-26 September, 2025 \\ Corfu, Greece }
\pacs{ 25.75.-q, 25.75.Gz, 25.75.Nq, 12.38.Mh} 
%\begin{document}
\maketitle
%\end{frontmatter}
\section{Introduction}
 One of the main goal of the study of relativistic heavy ion collisions is to study the deconfined matter, known as Quark-Gluon Plasma (QGP), which is expected to form at large densities. It has been suggested that the transition from hadronic to QGP state can be treated by the percolation theory \cite{celik}. The formulation of percolation problem is concerned with elementary geometrical objects placed on a random d-dimensional lattice. The objects have a well defined connectivity radius $\lambda$, and two objects can communicate if the distance between them is less than $\lambda$. Several objects can form a cluster of communication. At certain density of the objects a infinite cluster appears which spans the entire system. This is defined by the dimensionless percolation density parameter $\xi$ \cite{inch}.

In nuclear collisions there is indeed, as a function of parton density, a sudden onset of large scale color connection. There is a critical density at
which the elemental objects from one large cluster, loosing their independent existence. Percolation would correspond to the onset of color deconfinement and it may be prerequisite for any subsequent QGP formation.  

In this talk results are presented for the initial temperature and the energy density $\varepsilon$ in $pp$, Pb-Pb, and Xe-Xe collisions at LHC energies
 using published data by ALICE Collaboration.   
\begin{figure}[thbp]
  \centering  
  \includegraphics[width=0.8\textwidth,height=1.8in]{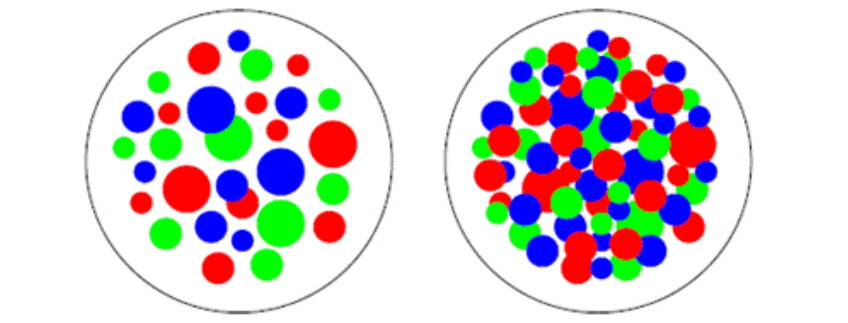}
% percolation picture 
%\vspace*{0.0cm}
\caption{Partonic cluster structure in the transverse collision plane at low (left) and (right) high parton density \cite{satzbook}.} 
\label{disk}
\end{figure}

\section{Clustering of Color Sources}   

Multiparticle production is currently described in terms of color strings stretched between the projectile and the target, which decay into new strings and subsequently hadronize to  produce the observed hadrons \cite{review15}. Color strings may be viewed as small areas in the transverse plane filled with color field created by colliding partons. In terms of gluon color field they can be considered as the color flux tubes stretched between the colliding partons. The mechanism of particle creation is the Schwinger $QED_{2}$ mechanism and is due to the color string breaking \cite{schw,wong}.

With growing energy and size of the colliding system, the number of strings grows, and they start to overlap, forming clusters, in the transverse plane very much similar to disks in two dimensional percolation theory as shown in 
Fig.~\ref{disk} \cite{inch,satzbook}. At a certain critical density, a macroscopic cluster appears that marks the percolation phase transition. This is the Color String percolation Model (CSPM) \cite{review15}.
The interaction between strings occurs when they overlap and the general result, due to the SU(3) random summation of charges, is a reduction in multiplicity and an increase in the string tension, hence an increase in the average transverse momentum squared, $\langle p_{T}^{2} \rangle$. 

We assume that a cluster of ${\it n}$ strings that occupies an area of $S_{n}$ behaves as a single color source with a higher color field $\vec{Q_{n}}$ corresponding to the vectorial sum of the color charges of each individual string $\vec{Q_{1}}$. The resulting color field covers the area of the cluster. As $\vec{Q_{n}} = \sum_{1}^{n}\vec{Q_{1}}$, and the individual string colors may be oriented in an arbitrary manner respective to each other , the average $\vec{Q_{1i}}\vec{Q_{1j}}$ is zero, and $\vec{Q_{n}^2} = n \vec{Q_{1}^2} $.

Knowing the color charge $\vec{Q_{n}}$ one can obtain the multiplicity $\mu$ and the mean transverse momentum squared $\langle p_{t}^{2} \rangle$ of the particles produced by a cluster of $\it n $ strings \cite{pajares2}
\begin{equation}
\mu_{n} = \sqrt {\frac {n S_{n}}{S_{1}}}\mu_{0};\hspace{5mm}
%\end{equation}
%\begin{linenomath*}
 %\begin{equation}                                      
\langle p_{t}^{2} \rangle = \sqrt {\frac {n S_{1}}{S_{n}}} {\langle p_{t}^{2} \rangle_{1}}
\label{mu}
\end{equation} 
%\end{linenomath*}
where $\mu_{0}$ and $\langle p_{t}^{2}\rangle_{1}$ are the mean multiplicity and $\langle p_{t}^{2} \rangle$ of particles produced from a single string with a transverse area $S_{1} = \pi r_{0}^2$. In the thermodynamic limit, one obtains an analytic expression \cite{pajares2,pajares1}
\begin{equation}
\langle \frac {n S_{1}}{S_{n}} \rangle = \frac {\xi}{1-e^{-\xi}}\equiv \frac {1}{F(\xi)^2};\hspace{5mm}
F(\xi) = \sqrt {\frac {1-e^{-\xi}}{\xi}}
\label{xi}
\end{equation}
where $F(\xi)$ is the color suppression factor. $\xi = \frac {N_{s} S_{1}}{S_{N}}$ is the percolation density parameter assumed to be finite when both the number of strings $N_{S}$ and total interaction area $S_{N}$ are large.  Eq.~(\ref{mu}) can be written as $\mu_{n}=F(\xi)\mu_{0}$ and 
$\langle p_{t}^{2}\rangle_{n} ={\langle p_{t}^{2} \rangle_{1}}/F(\xi)$.  
The critical cluster which spans $S_{N}$, appears for $\xi_{c} \ge$ 1.2 \cite{satz1}.
 
It is worth noting that CSPM is a saturation model similar to the Color Glass Condensate (CGC), where $ {\langle p_{t}^{2} \rangle_{1}}/F(\xi)$ plays the same role as the saturation momentum scale $Q_{s}^{2}$ in the CGC \cite{cgc,perx}. 
\section{ Color Suppression Factor $F(\xi)$}
\begin{figure}[thbp]
\centering        
\vspace*{-0.2cm}
\includegraphics[width=0.7\textwidth,height=4.5in]{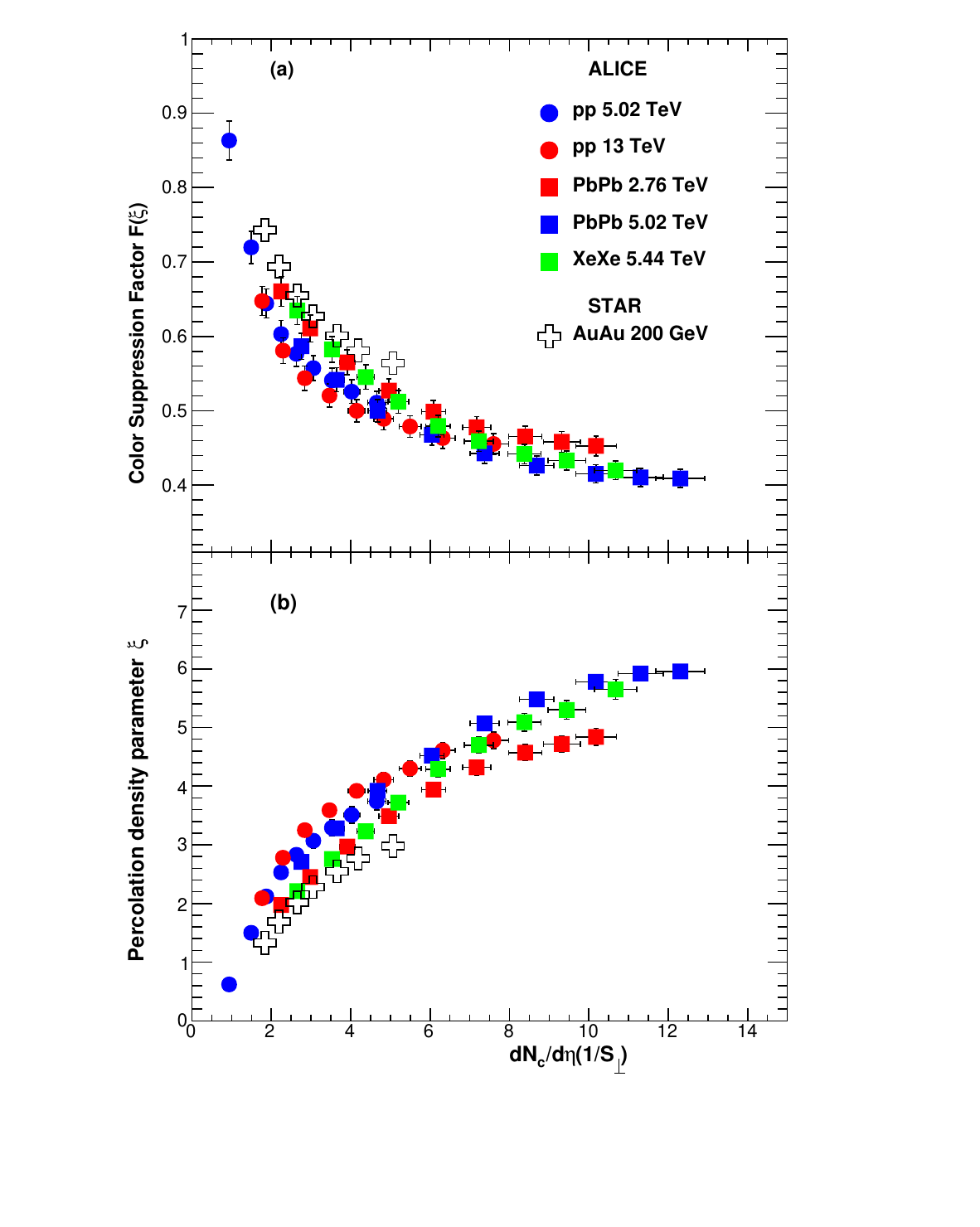}
\vspace*{-1.5cm}
\caption{Color Suppression Factor $F(\xi)$ in ${\it pp}$, Pb-Pb and Xe-Xe collisions vs $dN_{ch}/d\eta$ scaled by the transverse area $S_{\perp}$. For ${\it pp}$ collisions $S_{\perp}$ is multiplicity dependent as obtained from IP-Glasma model \cite{cross}. In case of Pb-Pb and Xe-Xe collisions the nuclear overlap area was obtained using the Glauber model \cite{glauber}.}
\label{fxintrackarea}
\end{figure}

In our earlier work $F(\xi)$ was obtained in Au-Au collisions by comparing the  charged hadron transverse momentum spectra from ${\it pp}$ and  Au-Au collisions\cite{review15}. To evaluate the initial value of $F(\xi)$ from data for Au-Au collisions, a parameterization of the experimental data of $p_{t}$ distribution in  ${\it pp}$ collisions  $\sqrt s$ = 200 GeV  was used \cite{review15}. The charged particle spectrum is described by a power law \cite{review15}

\begin{equation}
  d^{2}N_{c}/dp_{t}^{2} = a/(p_{0}+p_{t})^{\alpha},
  \label{spectra1}
\end{equation}
where $a$ is the normalization factor, $p_{0}$ and $\alpha$ are fitting parameters with $p_{0}$= 1.98 and  $\alpha$ = 12.87 \cite{review15}. This parameterization is used in high multiplicity ${\it pp}$ collisions to take into account the interactions of the strings \cite{review15}
\begin{equation} 
  \frac{d^{2}N_{c}}{dp_{T}^{2}} = \frac{a}{(p_{0} \sqrt {F(\xi)_{pp}/F(\xi)_{pp}^{mult}}+{p_{T}})^{\alpha}}.
  \label{spectra2} 
\end{equation}
where $F(\xi)_{pp}^{mult}$ is the multiplicity dependent color suppression factor.  In $pp$ collisions $F(\xi)_{pp} \sim$ 1 at low energies due to the low overlap probability.
In the present work we have extracted the color suppression factor $F(\xi)$ in high multiplicity events in ${\it pp}$ collisions using ALICE data from the transverse momentum spectra of charged particles at $\sqrt s$ =  5.02 and 13 TeV \cite{alicepp}.
For Pb-Pb at $\sqrt {s_{NN}} = $ 2.76 and 5.02 TeV and Xe-Xe at $\sqrt {s_{NN}} = $ 5.44 TeV $F(\xi)$ has been extracted from the spectra at various centralities \cite{alicepb,alicexe}.
The spectra were fitted using  Eq.~(\ref{spectra2}) in the softer sector with $p_{t}$ in the range 0.12-1.0 GeV/c.  
Figure~\ref{fxintrackarea} shows the Color Suppression Factor $F(\xi)$ in ${\it pp}$, $Pb-Pb$ and $Xe-Xe$ collisions as a function of charged particle multiplicity  $dN_{ch}/d\eta$ scaled by the transverse area $S_{\perp}$. For ${\it pp}$ collisions $S_{\perp}$ is multiplicity dependent as obtained from IP-Glasma model \cite{cross}. In case of $Pb-Pb$ and $Xe-Xe$ collisions the nuclear overlap area was obtained using the Glauber model \cite{glauber}. 

\section{Temperature}
The connection between $F(\xi)$ and the temperature $T(\xi)$ involves the Schwinger mechanism (SM) for particle production. 
The Schwinger distribution for massless particles is expressed in terms of $p_{t}^{2}$ \cite{wong}
\begin{equation}
dn/d{p_{t}^{2}} \sim \exp (-\pi p_{t}^{2}/x^{2})
\label{swing}
\end{equation}
where the average value of the string tension is  $\langle x^{2} \rangle$. The tension of the macroscopic cluster fluctuates around its mean value because the chromo-electric field is not constant.
The origin of the string fluctuation is related to the stochastic picture of 
the QCD vacuum. Since the average value of the color field strength must 
vanish, it cannot be constant but changes randomly from point to point \cite{bialas}. Such fluctuations lead to a Gaussian distribution of the string tension
\begin{equation}
  \frac{dn}{dp_{t}^{2}} \sim \sqrt \frac{2}{<x^{2}>}\int_{0}^{\infty}dx
  \exp (-\frac{x^{2}}{2<x^{2}>}) \exp (-\pi \frac{p_{t}^{2}}{x^{2}})
\end{equation}
 which gives rise to  thermal distribution \cite{bialas}
\begin{equation}
\frac{dn}{dp_{t}^{2}} \sim \exp (-p_{t} \sqrt {\frac {2\pi}{\langle x^{2} \rangle}} ),
\label{bia}
\end{equation}
with $\langle x^{2} \rangle$ = $\pi \langle p_{t}^{2} \rangle_{1}/F(\xi)$. 
The temperature is expressed as \cite{review15}  
\begin{equation}
T(\xi) =  {\sqrt {\frac {\langle p_{t}^{2}\rangle_{1}}{ 2 F(\xi)}}}.
\label{temp}
\end{equation}
\begin{figure}[thbp]
\centering        
\vspace*{1.0cm}
\includegraphics[width=0.7\textwidth,height=3.5in]{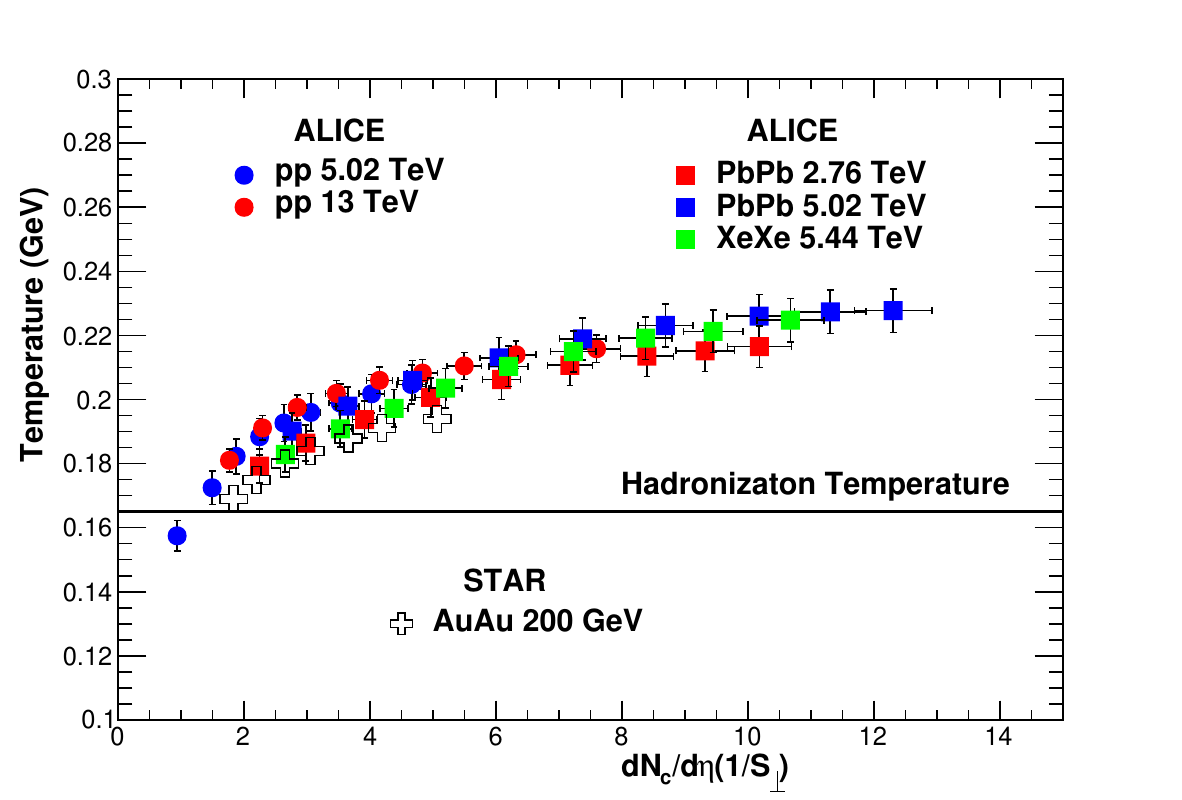}
\vspace*{-0.2cm}
\caption{ Temperature vs  $dN_{ch}/d\eta$ scaled by  $S_{\perp}$ from ${\it pp}$, Pb-Pb and Xe-Xe collisions. The horizontal  line  at $\sim 165 $ MeV is the universal hadronization temperature \cite{bec1}. } 
\label{tempfig}
\end{figure}
The string percolation density parameter $\xi$ which characterizes the percolation clusters measures the initial temperature of the system. Since this cluster 
covers most of the interaction area, the temperature becomes a global 
temperature determined by the string density. In this way at $\xi_{c}$ = 1.2 the connectivity percolation transition at $T(\xi_{c})$ models the thermal deconfinement transition.

Figure~\ref{tempfig} shows a plot of temperature as a function of
$dN_{ch}/d\eta$ scaled by $S_{\perp}$. The horizontal line at $\sim $ 165 MeV is the universal hadronization temperature obtained from the systematic comparison of the statistical model parametrization of hadron abundances measured in high energy $e^{+}e^{-}$, $pp$, and $AA$  collisions \cite{bec1}.
The temperatures obtained in higher multiplicity events are consistent with the creation of deconfined matter in $ pp $ collisions at  $\sqrt s$ = 5.02 and 13 TeV.

\begin{figure}[thbp]
\centering        
\vspace*{0.5cm}
\includegraphics[width=0.75\textwidth,height=4.0in]{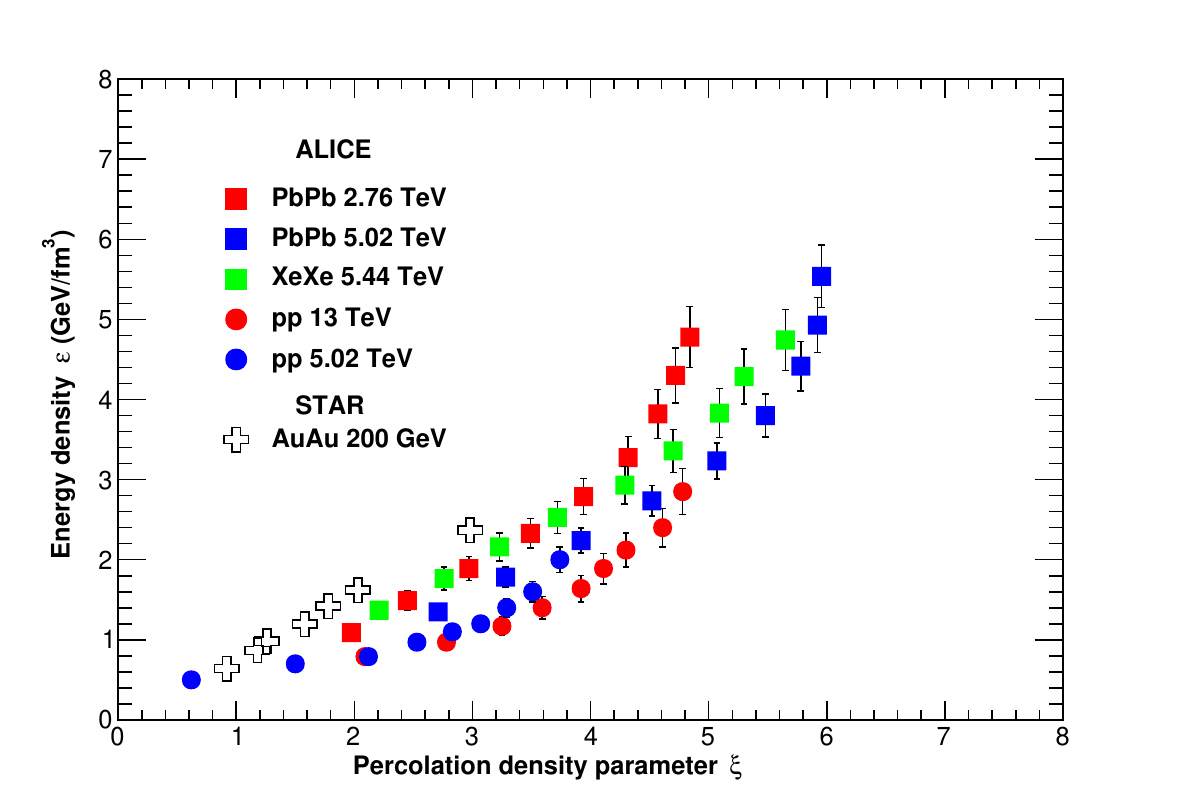}
\vspace*{-0.5cm}
\caption{Energy density ($\epsilon$) as a function 
of the percolation density parameter ($\xi$). Upper panel shows the pp collision
 data at $\sqrt s $ = 5.02 and 13 TeV. Lower panel shows the data for Pb-Pb at
 $\sqrt {s_{NN}}$ = 2.76 and 5.02 TeV and Xe-Xe data at $\sqrt {s_{NN}}$ = 5.44 
TeV. } 
\label{enxi}
\end{figure}
 In CSPM the strong color field inside the large cluster produces de-acceleration of the primary $q \bar q$ pair which can be seen as a thermal temperature by means of Hawking-Unruh effect. This implies that the radiation temperature is determined by the transverse extension of the color flux tube/cluster in terms of the string tension \cite{casto1}.
\begin{equation}
T =  {\sqrt \frac {\sigma}{2\pi}}
\label{haw}
\end{equation}

\section{Energy Density }
Among the most important and fundamental problems in finite-temperature QCD are the calculation of the bulk properties of hot QCD matter and characterization of the nature of the QCD phase transition. 
The QGP according to CSPM is born in local thermal equilibrium  because the temperature is determined at the string level. After the initial temperature $ T > T_{c}$ the  CSPM perfect fluid may expand according to Bjorken boost invariant 1D hydrodynamics \cite{bjorken}

\begin{equation}
\varepsilon = \frac {3}{2}\frac { {\frac {dN_{c}}{dy}}\langle m_{t}\rangle}{S_{n} \tau_{pro}}
\label{bjo}
\end{equation}
where $\varepsilon$ is the energy density, $S_{n}$ nuclear overlap area, and $\tau$ the proper time. 
\begin{equation}
%\tau_{pro} = \frac {2.405\hbar}{mc^{2}}.
\tau_{pro} = \frac {2.405\hbar}{\langle m_{t}\rangle}
\end{equation}
Above the critical temperature only massless particles are present in CSPM. 
 From the measured value of  $\xi$ and $\varepsilon$, as shown in Fig. \ref{enxi}, it is found  that $\varepsilon$ is proportional to $\xi$ for the range 
 $1.2 < \xi < 4.0 $. Above $\xi \sim $  4 the energy density $\varepsilon$ rises faster compare to $\xi < $ 4.

% et4
\begin{figure}[thbp]
\centering        
\vspace*{1.0cm}
\includegraphics[width=0.70\textwidth,height=3.5in]{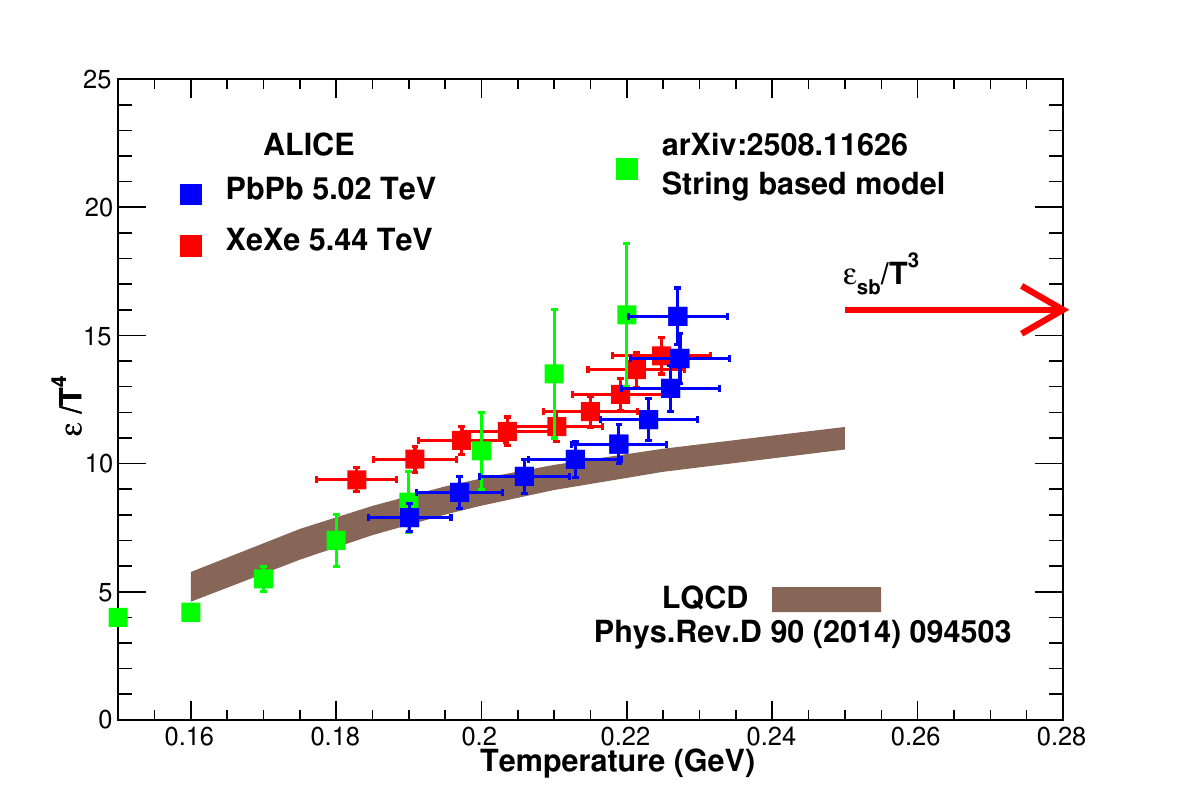}
\vspace*{-0.2cm}
\caption{Dimensionless quantity $\varepsilon/T^{4}$ as a function of temperature from CSPM and LQCD calculation from HotQCD Collaboration \cite{lattice14}. The CSPM values at higher temperature T > 200 MeV are obtained extrapolating from lower temperature.} 
\label{et4}
\end{figure}
Energy density has been obtained in lattice set up of (2+1)-flavor QCD using the HISQ action and the tree-level improved gauge action \cite{lattice14}.
Figure \ref{et4} shows dimensionless quantity $\varepsilon/T^{4}$ as a function of temperature both from CSPM and LQCD. It is observed that CSPM results differ from LQCD results above the temperature of T $\sim$  210 MeV in case of Pb-Pb and Xe-Xe collisions. In $pp$ collisions
CSPM results are lower as compared to LQCD simulations.  
Beyond T$\sim$ 210 MeV $ \varepsilon/T^{4}$ in CSPM rises much faster and reaches the ideal gas value of  $ \varepsilon/T^{4}$ $\sim$ 16 at T $\sim$ 230 MeV.
\section{Interpretation}
\subsection{Possible new phase of thermal QCD \cite{Ivan}}
%Andrei Alexandru and Ivan Horváth
%Phys. Rev. D 100, 094507 (2019)
%arXiv: 2305.09459
Recently, it has been pointed out using lattice simulations that in
addition of the standard crossover phase transition at T $\sim$ 155 MeV,
the existence of a new infrared phase transition at temperature T,
$200 < T_{IR} < 250$ MeV. In this phase, asymptotic freedom works and
therefore there is no interaction. In between these two temperatures
there is possible coexistence of the short and long-distance scales
\cite{Ivan}.
\subsection{Three regimes of QCD \cite{Glozman}}
% L. Ya. Glozman Int. J. Mod. Phys. A 36 (2021) 2044031
In addition to the hadron gas and the quark gluon plasma phases, it was
proposed the existence of a stringy fluid phase based on the chiral spin
symmetry. At temperatures below the pseudocritical temperature $T_{c} \sim $ 155
MeV the QCD matter is hadron gas with spontaneously broken chiral
symmetry. Within the window $T_{c} -3T_{c}$ the hot QCD is represented by stringy
fluid with restored chiral and approximate chiral spin symmetries.
Above $\sim 3T_{c}$ the chiral symmetry disappears and one observes a
smooth transition to partonic degrees of freedom, i.e. to a
quark-gluon plasma \cite{Glozman}.
 \begin{figure}[thbp]
\centering        
\vspace*{1.0cm}
\includegraphics[width=0.75\textwidth,height=3.5in]{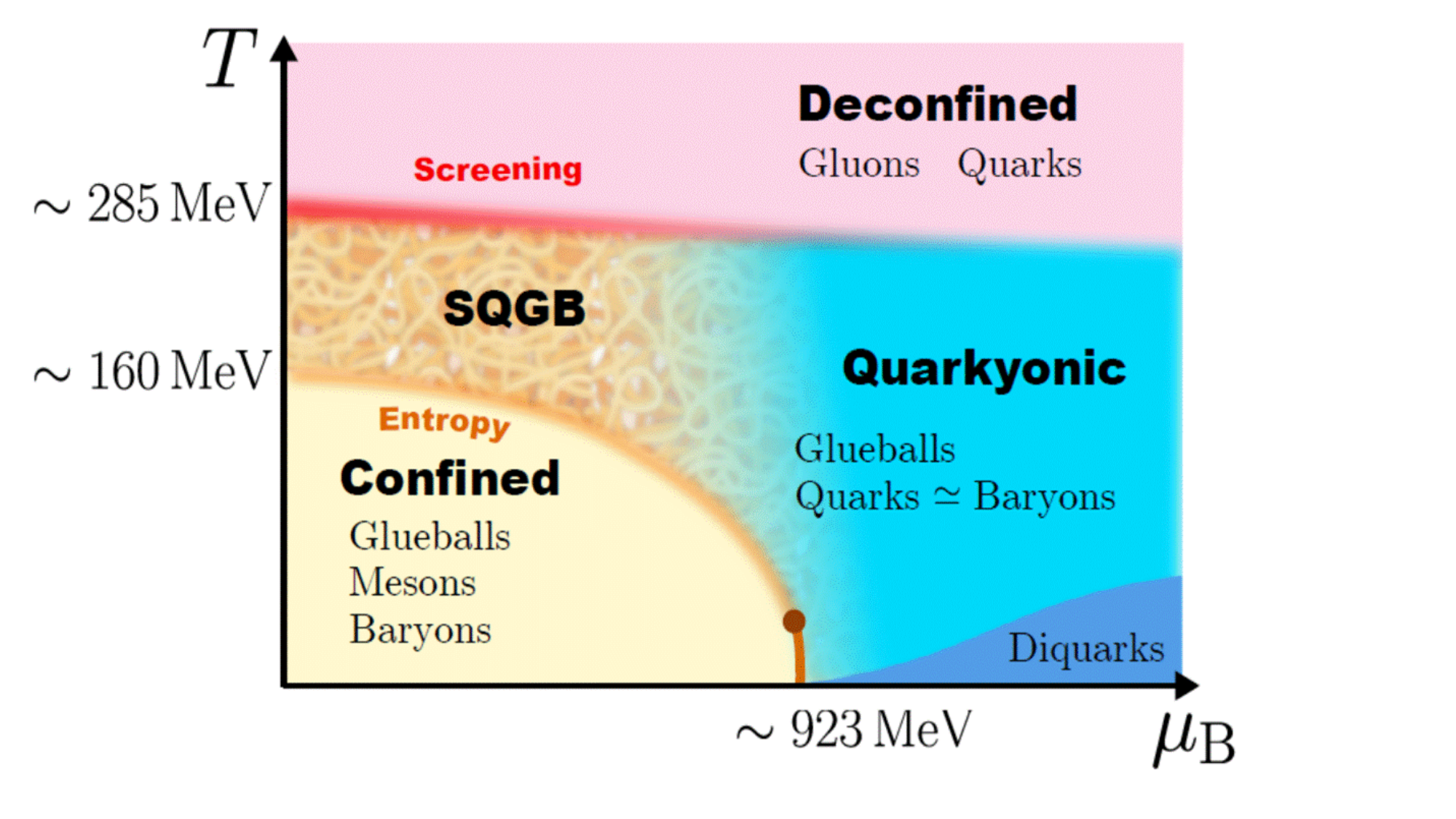}
\vspace*{-0.2cm}
\caption{New and realistic phase diagram for $N_{c}$ =3 including a window of the SQGB regime \cite{sqgp1}}.
\label{sqgb}
\end{figure}
\subsection{Center vortex evidence for a second finite-
  temperature QCD transition \cite{Mickley}}
% Mickley et al arXiv: 2411.19446
Evidence for the existence of a second finite-temperature transition
in quantum chromodynamics (QCD) is obtained through the study of
vortex geometry and its evolution with temperature.
This transition is a deconfinement transition associated with the loss
of vortex percolation at temperature $T_{d}$= 321 MeV \cite{Mickley}.
\subsection{A new perspective on thermal transition in QCD
  \cite{Hanada}}
% Hanada et al arXiv:2310.01940
Another proposal is the partial deconfinement which implies an intermediate
phase in which color degree of freedom split into the confined and deconfined
sectors. The partially deconfined phase is dual to the small black hole
that lies between large black hole and the graviton gas. The transition
to partial deconfinement would corresponds to the standard cross over.
From partial to complete deconfinement would be accompanied by additional transition of the energy density. The three phases are characterized by the behavior of the function embedded in the phases of the Polyakov line \cite{Hanada}.
\subsection{A New State of Matter between the Hadronic Phase and the Quark-Gluon Plasma? \cite{sqgp1,sqgp2}}
% Fujimoto et al arXiv:2506.00237, arXiV:2508.11626
Lattice-QCD simulations and theoretical arguments hint at the
existence of an intermediate phase of strongly interacting matter
between a confined hadron gas and a deconfined Quark-Gluon Plasma
(QGP). We qualitatively and semi-quantitatively explore and
differentiate the phase structures in the temperature window from the
QCD pseudo-critical temperature $T_{c} \simeq $160 MeV to the pure-gluonic
deconfinement temperature $T_{d} \simeq $ 285 MeV.
We have argued that the phases of QCD at finite temperature and
density might include a new intermediate phase, which is named as
Spaghetti of Quarks with Gluballs, the SQGB \cite{sqgp1}. Figure ~\ref{sqgb}
shows the plausible picture of new phase diagram. 

In the presence of the SQGB, three phases can be identified:
At low temperature ($T < T_{c}$), there is a gas of mesons, baryons, and
possibly glueballs.
In the intermediate SQGB phase ($T_{c} < T < T_{d}$), the quarks appear as
thermal degrees of freedom despite being at least partially confined
At high temperature ($T > T_{d}$), the system is fully deconfined in the form of
QGP \cite{sqgp2}. The SQGB simulation for $ \varepsilon/T^{4}$ is shown in
Fig.~\ref{et4} along with CSPM and LQCD results. It is worth mentioning that
SQGB results also shows a sudden rise $T\sim 200$ as seen in our result.

 \section{Conclusions}
  We have used the Color String Percolation Model to explore the initial stage in ${\it pp}$, Pb-Pb, and Xe-Xe collisions at LHC energies and determined the thermalized initial temperature of the hot nuclear matter at an initial time $\sim$ 1 $fm/c$. For the first time both the temperature and the energy density of the hot nuclear matter have been obtained from the measured charged particle spectra using ALICE data for ${\it pp}$ collisions at $\sqrt {s}$ = 5.02 and 13 TeV along with Pb-Pb at $\sqrt {s_{NN}}$ = 2.76 and 5.02 TeV and 
  Xe-Xe at $\sqrt {s_{NN}}$ = 5.44 TeV.

  Our results for Pb-Pb and Xe-Xe collisions show a sharp increase in
  $ \varepsilon/T^{4}$ above T $\sim$ 210 MeV and reaching the ideal gas of quarks and gluons value of $ \varepsilon/T^{4} \sim$ 16 at temperature $\sim $ 230 MeV.
 The results indicate that there are two temperatures. The first one T $\sim$ 160 MeV would correspond to the restoration of chiral symmetry and partial deconfinement. The second transition 
T $\sim$ 220 MeV leads to a total deconfined phase. At this temperature there
is a transition from the fluid behavior of QCD matter strongly interacting
to a quasi free gas of quarks and gluons. This results is consistent with the
 recently published string based SQGB work. 
\section{Acknowledgment}
 C.P. thanks the grant Maria de Maeztu Unit of excelence MDM-2016-0682 of Spain, the support of Xunta de Galicia under the project  ED431C 2017 and  project FPA 2017-83814 of Ministerio de Ciencia e Innovacion of Spain and FEDER.

\end{document}